\documentclass[12pt]{iopart}
\usepackage{graphics}
\usepackage{graphicx}

\newcommand{\nsp}{\vspace*{-3mm}}
\newcommand{\order}{{\mathcal O}}
\newcommand{\bJ}{{\mbox{\boldmath $J$}}}
\newcommand{\bra}{\langle}
\newcommand{\ket}{\rangle}

\begin{document}
\letter{Dynamics of on-line Hebbian learning
with structurally unrealizable restricted training sets}
\author{Jun-ichi Inoue \dag
\footnote[3]{j\underline{\,\,\,}inoue@complex.eng.hokudai.ac.jp}
and A.C.C. Coolen \ddag}

\address{\dag\, Complex Systems Engineering, Graduate
School of Engineering, Hokkaido University,
N13-W8, Kita-ku, Sapporo 8628, Japan}

\address{\ddag\, Department of Mathematics, King's
College London, The Strand, London WC2R 2LS, UK}


\begin{abstract}
We present an exact solution for the dynamics of on-line Hebbian learning in
neural networks, with restricted and unrealizable training sets.
In contrast to other studies on learning with restricted training sets,
unrealizability is here caused by structural mismatch,
rather than data noise: the teacher
 machine is a perceptron with a reversed wedge-type transfer function,
while the student machine is a perceptron
with a sigmoidal transfer function.
We calculate the glassy dynamics of the macroscopic performance measures,
training error and generalization error, and the
(non-Gaussian) student field distribution. Our
results, which find excellent confirmation in numerical simulations,
provide a new benchmark test for general formalisms with which to study
unrealizable learning processes with restricted training sets.
\end{abstract}
On-line learning processes
in artificial neural networks have been studied using
statistical mechanical techniques for about a decade now (see e.g. \cite{OK,MC} for reviews).
Initially, most dynamical studies were restricted to
the regime where the number of training examples
is larger than the number of learning steps, since this
generally leads to Gaussian field distributions
and relatively simple non-glassy dynamics.
In practical situations, however, it is usually
difficult to acquire large training sets, and one is therefore
often forced to recycle the data in the
training set.
The latter situation,
characterized by the presence of disorder (the composition of the training set) and non-trivial
dynamics, was studied in e.g. \cite{Horner1,Horner2}
(for binary weights), and in \cite{CS1,CS2,WongLi,CM,CSX,HC} (for continuous
weights). These studies generally involve approximations at some
stage. This motivated \cite{Rae}, where it was shown how for the special case of
on-line Hebbian learning the dynamics can be solved exactly (for restricted training sets),
providing an excellent
benchmark for general theories and approximation schemes.
Some of the studies mentioned above involved learning from restricted but
unrealizable training sets, where it is impossible for the student
to achieve perfect performance,
even if an infinitely large training set had been available.
This could result from corruption by noise of realizable data (as in e.g.
\cite{CM,Rae}), or from structural
mismatch between teacher and student.
A typical toy model to realize the latter situation is obtained by using
a perceptron with a reversed wedge transfer function
as a teacher machine to train an ordinary perceptron
\cite{Watkin,inoue,Kabashima} (note: there is also a relation between simple
perceptrons with reversed wedge transfer functions and the so-called parity machines).
Since all dynamical studies with restricted but unrealizable
training sets have so far been carried out only for the data noise scenario,
it would be of considerable interest to investigate exactly solvable
models with restricted training sets but unrealizability due to structural mismatch.
In this letter, we carry out such a study: we solve the dynamics of on-line
Hebbian learning from unrealizable restricted training sets, for a
teacher-student scenario where teacher and student have different transfer functions
(a reverse-wedge and a sigmoidal one, respectively).

We investigate on-line learning in a ordinary student perceptron
$S$ (whose weight vector is
denoted by $\mbox{\boldmath $J$}$),
which tries to learn a task defined by a teacher
perceptron $T_{a}$ (whose weight
vector is denoted by $\mbox{\boldmath $B$}$). The teacher is equipped with
a reversed wedge transfer function, i.e. $T_{a}(y)={\rm sgn}[y(a-y)(a+y)]$
where $y=\mbox{\boldmath $B$} \,\cdot\, \mbox{\boldmath $\xi$}$
and $\mbox{\boldmath $\xi$} \in \{-1,1\}^{N}$ is the
input vector, whereas $S(x)={\rm sgn}[x]$ with $x=\mbox{\boldmath $J$} \,\cdot\, \mbox{\boldmath
$\xi$}$.
The teacher's weight vector $\mbox{\boldmath $B$}$
is normalized
such that $\mbox{\boldmath $B$}^{2}=1$, with
$B_{i}=\order(N^{-1/2})$ for each $i$.
It is clear that in the limits $a\to 0$ and $a\to\infty$
(where $a$ characterizes the width of the reverse wedge)
 the task becomes realizable for the
student, since $T_{0}(y)={\rm sgn}[-y]$ and $T_{\infty}(y)={\rm sgn}[y]$.

We define the conventional order parameters
$Q[\bJ]\,\equiv\,\mbox{\boldmath $J$}^{2}$ and
$R[\bJ]\,\equiv\, \mbox{\boldmath $B$} \cdot
\mbox{\boldmath $J$}$.
One of the main quantities of interest is
the generalization error $E_g$, the probability
of disagreement between teacher and student for input vectors  taken
randomly from the {\em full} set of all possible inputs:
\begin{equation}
E_{g} \equiv  \langle  {\Theta}[-T_{a}(y) S(x)] \rangle_{\mbox{\boldmath
$\xi$}}
=\int_{0}^{a}\!Du~{\rm erf}[r^{+}(u)]
+\int_{a}^{\infty}\!Du~{\rm erf}[r^{-}(u)],
\label{gene}
\end{equation}
where $r^{\pm}(u) \,\equiv\, \pm Ru/\sqrt{2(Q\!-\!R^{2})}$,
$Du\,\equiv (2\pi)^{-\frac{1}{2}}du~ {\rm e}^{-u^{2}/2}$,
$\Theta[\cdots]$ is the step
function,
and
$\langle \cdots \rangle_{\mbox{\boldmath $\xi$}}$
denotes averaging over all $\mbox{\boldmath $\xi$}\in\{-1,1\}^N$.
It was shown in  \cite{inoue} that
the optimal normalized overlap $r=R/\sqrt{Q}$ (giving the
smallest value of the
generalization error) equals $1$ as long as
the reversed wedge parameter $a$ is greater than
$a=a_{\rm c1}=0.8$; $r$ suddenly
changes from $1$ to $r_{*}=
-\sqrt{(2 \log 2-a^{2})/2\log 2}$ at $a=a_{\rm c1}$.

For this model system, we use the following
on-line Hebbian learning rule
\begin{equation}
\mbox{\boldmath $J$}(\ell+1)=\left(
1-\frac{\gamma}{N}
\right)\mbox{\boldmath $J$}(\ell)+
\frac{\eta}{N} T_{a}^{\mu (\ell)} \mbox{\boldmath $\xi$}^{\mu(\ell)}
\label{updateJ}
\end{equation}
where $\ell$ indicates the learning step, and $\eta$ and $\gamma$ represent
the learning rate and the weight decay, respectively.
The student learns from data picked randomly from the restricted training set
$D=\{(\mbox{\boldmath $\xi$}^{\mu},T_{a}^{\mu}),~
\mu=1,\ldots,p = N\alpha\}$.

To calculate macroscopic physical observables, averaged  over the disorder (the composition of the
training set) at any time, we need to distinguish between
 two averaging procedures \cite{Rae}.
The first is the average over all possible `paths'
${\Omega}=\{\mu(0),\mu(1),\cdots, \mu(\ell), \cdots \}$ defining the
actual sampling order from the training set:
\begin{eqnarray}
\langle f(\mbox{\boldmath $\xi$}^{\mu(l)},
T_{a}) \rangle_{\Omega} &  \equiv  & \frac{1}{p}
\sum_{\mu=1}^{p}f(\mbox{\boldmath $\xi$}^{\mu}, T_{a})
\label{pass_ave}
\end{eqnarray}
The second is averaging over all training sets:
\begin{eqnarray}
\langle f[
(\mbox{\boldmath $\xi$}^{1},T_{a}^{1}),{\cdots},
(\mbox{\boldmath $\xi$}^{p},T_{a}^{p})
]\rangle_{\rm sets} & \equiv &
2^{-pN}
\sum_{\mbox{\boldmath $\xi$}^{1}}
{\cdots}
\sum_{\mbox{\boldmath $\xi$}^{p}}
\sum_{T_{a}^{1},{\cdots}, T_{a}^{p}}
\prod_{\mu}
P(T_{a}^{\mu}|\mbox{\boldmath $\xi$}^{\mu})  \nonumber \\
\mbox{} & &
\times~f[
(\mbox{\boldmath $\xi$}^{1}, T_{a}^{1}),
\cdots,
(\mbox{\boldmath $\xi$}^{p}, T_{a}^{p})]
\label{sets_ave}
\end{eqnarray}
The key to the full solvability of the present model, as in
\cite{Rae}, is the fact that \eref{updateJ} allows us to
write $\mbox{\boldmath $J$}(m)$
(the student's weight vector
at $m$-th step) in explicit form as
\begin{equation}
\mbox{\boldmath $J$}(m) =
\sigma^{m} \mbox{\boldmath $J$}(0)
+\frac{\eta}{N}
\sum_{\ell=0}^{m-1}
\sigma^{m-\ell-1}
T_{a}^{\mu (\ell)}
\mbox{\boldmath $\xi$}^{\mu (\ell)}
\label{eq:explicit}
\end{equation}
where $\sigma \equiv (1-\eta/N)$. The above
averaging procedures can now be carried out exactly.

\begin{figure}[t]
\begin{center}
\includegraphics[width=.44\linewidth,height=0.44\linewidth]{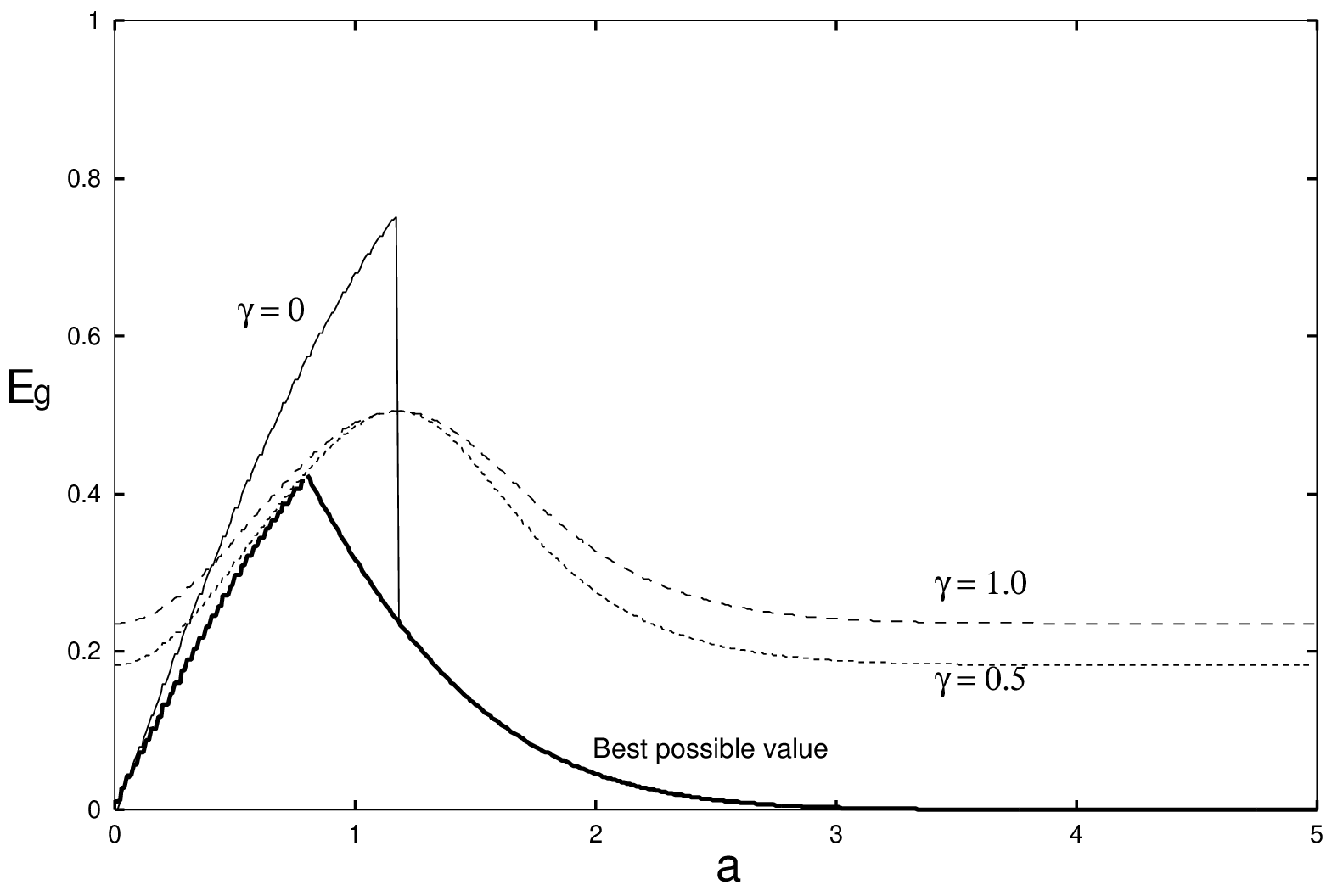}
\hspace*{4mm}
\includegraphics[width=.44\linewidth,height=0.44\linewidth]{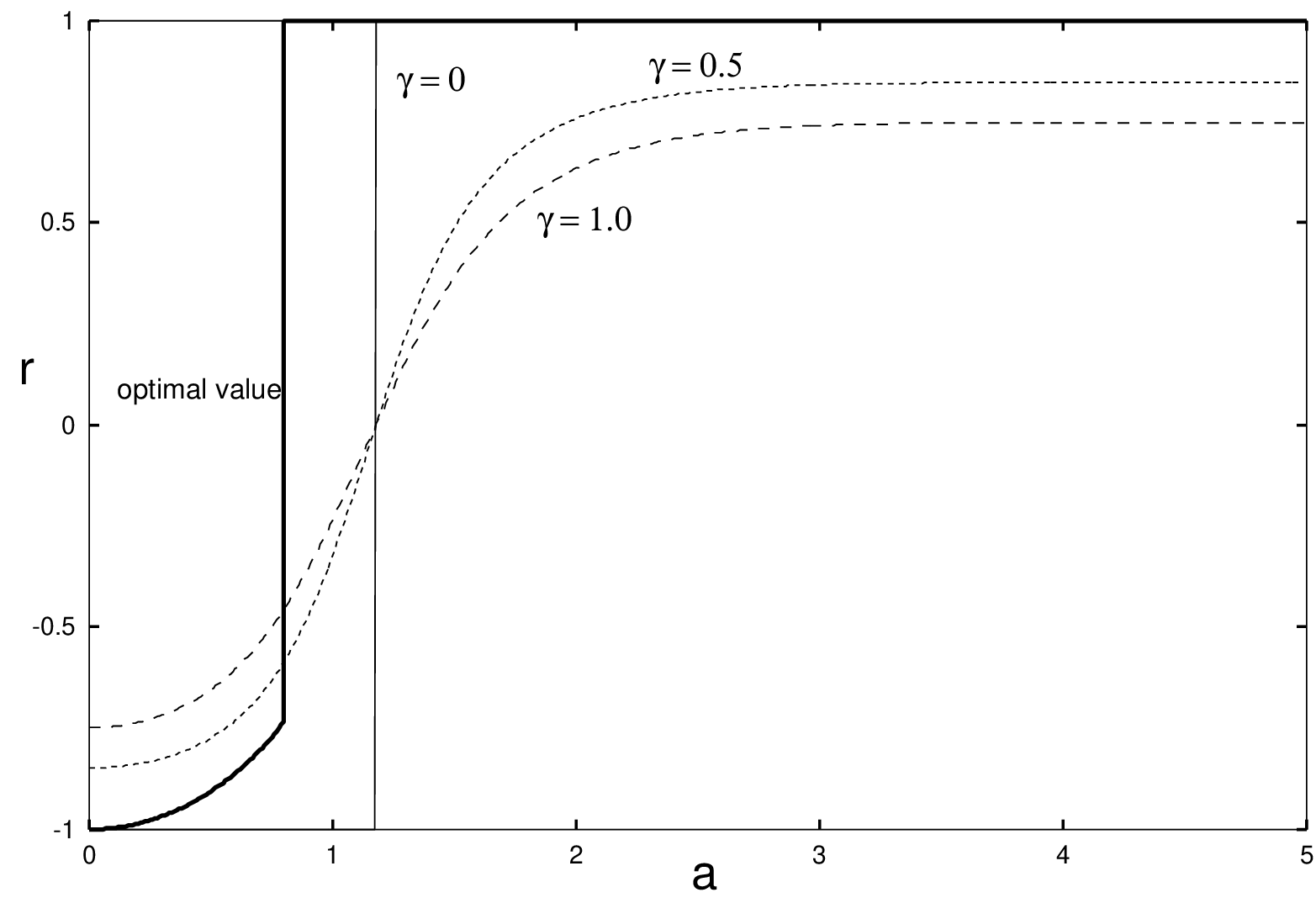}
\end{center}
\nsp
\caption{
Left: The asymptotic generalization error
as a function of the width of the
reversed wedge $a$
in the limit of $\alpha \rightarrow \infty$,
for $\gamma=0$ (\full), $\gamma=0.5$ (\dotted) and $\gamma=1$ (\dashed).
We  chose $\eta=1$.
Right: The corresponding normalized overlap $r=R/\sqrt{Q}$ which gives
the generalization error in the left figure.
The best possible values for the generalization error
and the optimal normalized overlap are shown by thick lines.
}
\label{fig1}
\end{figure}

In order to evaluate the training time dependence of
the generalization error \eref{gene},
following \cite{Rae}, we first
calculate the following two macroscopic observables
\begin{equation}
Q(t) \equiv
\displaystyle{\lim_{N{\rightarrow}\infty}}
\langle \langle Q[\bJ(m)] \rangle_{\Omega} \rangle_{\rm sets},
~~~~~~~~~
R(t) \equiv \displaystyle{\lim_{N{\rightarrow}\infty}}
\langle \langle  R[\bJ(m)] \rangle_{\Omega} \rangle_{\rm sets},
\label{ave_QR}
\end{equation}
where $t \equiv m/N$.
Squaring (\ref{eq:explicit}) gives
\begin{eqnarray*}
\langle \langle Q[\bJ(m)] \rangle_{\Omega} \rangle_{\rm sets} & = &
{\sigma}^{2Nt}Q_{0}+
\frac{2\eta}{N}
\sum_{\ell=0}^{m-1}
{\sigma}^{2m-\ell-1}
\langle \langle
\mbox{\boldmath $J$}_{0} \cdot \mbox{\boldmath $\xi$}^{\mu(\ell)}
T_{a}^{\mu(\ell)}\rangle_{\Omega}
 \rangle_{\rm sets} \nonumber \\
\mbox{} & + &
\frac{\eta^{2}}{N^{2}}
\sum_{\ell,\ell^{\prime}=0}^{m-1}
\langle \langle
{\sigma}^{m-\ell-1}{\sigma}^{m-\ell^{\prime}-1}
\mbox{\boldmath $\xi$}^{\mu(\ell)} \cdot \mbox{\boldmath $\xi$}^{\mu(\ell^{\prime})}
T_{a}^{\mu(\ell)}T_{a}^{\mu(\ell^{\prime})}
\rangle_{\Omega} \rangle_{\rm sets}.
\end{eqnarray*}
After calculating the averages
$\langle \cdots \rangle_{\Omega}$ and
$\langle \cdots \rangle_{\rm sets}$, and taking $N\to\infty$, we then obtain
\begin{eqnarray}
Q(t) & = &
{\rm e}^{-2\gamma t}Q_{0}
+\frac{2\eta {\rho}_{a}R_{0}}{\gamma}
{\rm e}^{-\gamma t}(1-{\rm e}^{-\gamma t})
+\frac{\eta^{2}}{2\gamma}
(1-{\rm e}^{-2\gamma t}) \nonumber \\
\mbox{} & & +
\frac{\eta^{2}}{\gamma^{2}}
(1/\alpha+{\rho}_{a}^{2})
(1-{\rm e}^{-\gamma t})^{2}
\label{Qt}
\end{eqnarray}
where we defined $\rho_a \equiv
\langle (\mbox{\boldmath $v$}
\cdot \mbox{\boldmath $\xi$})
T_a (\mbox{\boldmath $B$} \cdot
\mbox{\boldmath $\xi$})
\rangle_{\mbox{\boldmath $\xi$}}=
\sqrt{2/\pi}(1\!-\!2 \,{\rm e}^{-a^{2}/2})$.
The quantity $\rho_{a}$
represents a kind of effective noise induced by the
reversed wedge of the teacher.
In a similar manner we obtain an exact expression for
the student-teacher overlap $R(t)$:
\begin{equation}
R(t)=
{\rm e}^{-\gamma t}R_{0}
+\frac{\eta \rho_{a}}{\gamma}
(1-{\rm e}^{-\gamma t}).
\label{Rt}
\end{equation}
The length of the component of $\mbox{\boldmath $J$}$ which is
orthogonal to $\mbox{\boldmath $B$}$, $\sqrt{Q-R^2}$,
is seen to remain independent of $\rho_a$.
This is easily understood.
The components of the
input vectors which are orthogonal to $\mbox{\boldmath $B$}$
are uncorrelated with the training outputs, so their evolution is not modified
by the effect of the reversed wedge.
From (\ref{Qt},\ref{Rt}), in turn, we immediately obtain the
generalization error at any time, via (\ref{gene}).
For $t\to\infty$ this becomes
\begin{equation}
\displaystyle{\lim_{t\rightarrow \infty}}
E_{g} =
\int_{0}^{a}\! Du~{\rm erf}[r_{*}^{+}(u)]+
\int_{a}^{\infty}\! Du~{\rm erf}[r_{*}^{-}(u)]
\label{asymge}
\end{equation}
with $r_{*}^{\pm}(u) \equiv \pm \rho_a u/\sqrt{\gamma +2/\alpha}$.
In \Fref{fig1},
we show the asymptotic value of $E_{g}$ for
$\alpha \rightarrow \infty$ (where we recover the unrestricted
training sets behaviour),
for different $\gamma$.
We see that for $\gamma=0$,
$E_{g}$ converges to $2\,{\rm erf}(a/\sqrt{2})$
for $a < a_{\rm c2} =\sqrt{2\log 2}$
and to $1-2\,{\rm erf}(a/\sqrt{2})$ for $a > a_{\rm c2}$,
with an asymptotic scaling form $E_g\sim\alpha^{-1/2}$ as $\alpha\to \infty$ \cite{inoue}.
On the other hand,
for $\gamma > 0$, $E_{g}$ converges to
$E_{g}^{*}|_{\alpha=\infty}$ as $E_g\sim\alpha^{-1}$.
For $\gamma =0$ and $a < a_{\rm c2}$,
the asymptotic generalization error $E_{g}$ is seen to be
{\em larger} than that corresponding to random guessing (over-training) \cite{inoue}.
When we introduce weight decay
this phenomenon disappears.
An optimal weight decay,
minimizing the asymptotic $E_{g}$, exists for $a <a_{\rm c2}$
and is given by $\gamma_{\rm opt} =
2a^{2}\rho_{a}^{2}/(2\log 2 -a^{2})$.

For finite $\alpha$ and short times, $t \ll 1/\gamma$,
we can expand \eref{Qt} and \eref{Rt}
with respect to $\gamma t$ and find
$R(t)= R_{0}+\eta \rho_a t,~
Q(t)-R^2(t)=Q_{0}-R_{0}^{2}
+\eta^{2}t+\eta^{2}t^{2}/\alpha$.
In this regime
the training time is too short for weight decay to have an effect.
For $t  \gg 1/\gamma$, on the other hand,
it is clear from \eref{Qt} and \eref{Rt}
that the order parameters $Q, R$
decay to their asymptotic values exponentially.
For the case of $\gamma \rightarrow 0$,
the small $\gamma t$ expansions are
valid for all time. Upon expanding $E_{g}$ with respect to $\gamma$
we obtain
\begin{eqnarray*}
E_{g}  \simeq
\int_{0}^{a}\!Dx~{\rm erf}\left[
\sqrt{\frac{\alpha}{2}}
\rho_a x
\right]
+\int_{a}^{\infty}\!Dx~ {\rm erf}
\left[
-\sqrt{\frac{\alpha}{2}}\rho_a x
\right] +
\frac{\alpha}{2t\sqrt{2\pi (1+\alpha \rho_a^{2})}}
\end{eqnarray*}

We next turn to the student field distribution $P_t(x)$.
If the number of examples in the training set is much larger than
the number of training steps (i.e. for $\alpha\to \infty$),
the student fields $x=\mbox{\boldmath $J$} \cdot
\mbox{\boldmath $\xi$}$ are described by a Gaussian distribution, due to
the central limit theorem.
For $\alpha<\infty$ however, where the training sets are restricted and
questions are recycled during
the training process,
complicated correlations build up
and the field distribution generally acquires a non-Gaussian shape.
In order to determine $P_{t}(x)$, we first
calculate the joint distribution for student fields, teacher
fields, and outputs (with $x,y\in\Re$ and $T_a\in\{-1,1\}$):
\begin{eqnarray*}
P(x,y,T_{a}) =\lim_{N\to\infty}
\frac{1}{p}\sum_{\mu=1}^{p}
\bra\bra~\delta (x-\mbox{\boldmath $J$} \cdot
\mbox{\boldmath $\xi$}^{\mu})
\delta (y - \mbox{\boldmath $B$} \cdot
\mbox{\boldmath $\xi$}^{\mu})\,
\delta_{T_{a},T_{a}^{\mu}}~\ket_{\Omega}\ket_{\rm sets},
\end{eqnarray*}
Its characteristic function is
\begin{eqnarray*}
\hat{P}_{t}(\hat{x},\hat{y},\hat{T})=
\langle {\rm e}^{-i(\hat{x}x+\hat{y}y+\hat{T}T)}
\rangle_{P_{t}(.,.,.)}
\end{eqnarray*}
where $\langle f(x,y,T_a) \rangle_{P_{t}(.,.,.)}=\int\!dxdy\sum_{T_a=\pm 1} P_{t}(x,y,T_{a})f(x,y,T_a)$.
Working out the (\ref{pass_ave},\ref{sets_ave}) averages we obtain
\begin{eqnarray*}
\hat{P}_{t}(\hat{x},\hat{y},\hat{T})& =&
\displaystyle{\lim_{N \rightarrow \infty}}
{\biggr \langle}
\frac{1}{p}
\sum_{\mu=0}^{p}
{\exp}[
-i\hat{x}\sigma^{Nt}
\mbox{\boldmath $J$}_{0} \cdot \mbox{\boldmath $\xi$}^{\mu}
-i\hat{y} \mbox{\boldmath $B$} \cdot
\mbox{\boldmath $\xi$}^{\mu}
-i\hat{T} T_{a}^{\mu}] \nonumber \\
\mbox{}&& \times
{\biggr \langle}
{\exp}\left(
-\frac{i\eta \hat{x}}
{N}\sum_{\ell=0}^{Nt}
\sigma^{Nt-\ell}
\mbox{\boldmath $\xi$}^\mu \cdot
\mbox{\boldmath $\xi$}^{\mu (\ell)}
T_{a}^{\mu (\ell)}
\right)
{\biggr \rangle}_{\Omega} {\biggr \rangle}_{\rm sets}.
\end{eqnarray*}
By using the general relation
\begin{eqnarray}
\hat{P}_{t}(\hat{x},\hat{y},\hat{T})=
\int\! dxdy \sum_{T_{a}=\pm 1}
e^{-i(\hat{x}x+\hat{y}y +\hat{T}T_{a})}
 P_{t}(x|y,T_{a})P(y,T_{a})
\label{relationP}
\end{eqnarray}
and some further algebra, following closely the procedure outlined in \cite{Rae} (to which we refer
for details), we then obtain
the probability density
$P_{t}(x)$ as
\begin{eqnarray}
P_{t}(x)&=&\int\! Dy \sum_{T_{a}=\pm 1}
P_{t}(x|y,T_{a})P(T_{a}|y) \nonumber \\
&=&
\int\! \frac{d\hat{x}}{2\pi}
{\rm e}^{-\frac{Q}{2}x^{2}+{\chi}_{\rm r}(\hat{x})}
\cos(\hat{x}x)\left[
{\cos}(\chi_{\rm i}(\hat{x}))
+G(\hat{x}R){\sin}(\chi_{\rm i}(\hat{x}))
\right] \nonumber \\
 & -&
4 {\rm e}^{-\frac{a^2}{2}}
\!\int\! \frac{d\hat{x}}{2\pi}
{\rm e}^{-\frac{Q}{2}\hat{x}^{2}
+{\chi}_{\rm r}(\hat{x})}
\cos(\hat{x}x) \sin
(\chi_{\rm i}(\hat{x}))
\int_{0}^{x\hat{R}}\!\!\!
\frac{d\lambda}{\sqrt{2\pi}}
{\rm e}^{\frac{\lambda^{2}}{2}}
\cos(a\lambda)
\label{Pt}
\end{eqnarray}
where we defined the functions $\chi_{\rm r}$,
$\chi_{\rm i}$ and $G$ as
\begin{eqnarray*}
\chi_{\rm r}(\hat{x}) \equiv
\frac{1}{\alpha}
\int_{0}^{t}\!
ds \left\{
\cos \,[\eta \hat{x} {\rm e}^{-\gamma (t-s)}]
-1
\right\},~~~~
\chi_{\rm i}(\hat{x}) \equiv
-\frac{1}{\alpha}
\int_{0}^{t}\!
ds \sin \,[\eta \hat{x} {\rm e}^{-\gamma (t-s)}]
\end{eqnarray*}
\begin{eqnarray*}
G(\Lambda) \equiv
{\rm e}^{\frac{1}{2}\Lambda^2}
\int\! Dz~\sin(\Lambda|z|) =
2 \int_{0}^{\Lambda}\!
\frac{d\lambda}{\sqrt{2\pi}}
{\rm e}^{\frac{\lambda^2}{2}}.
\end{eqnarray*}

\begin{figure}[t]
\begin{center}
\includegraphics[width=0.23\linewidth,height=0.43\linewidth]{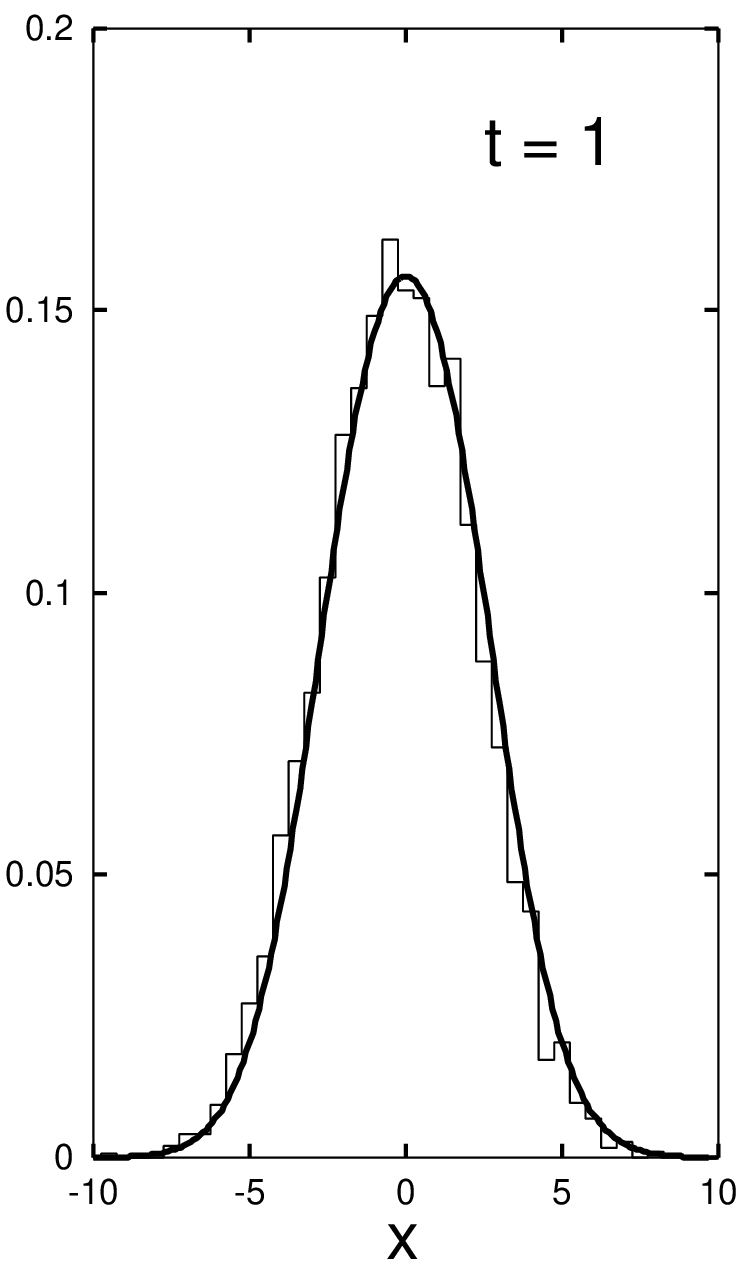}
\includegraphics[width=0.23\linewidth,height=0.43\linewidth]{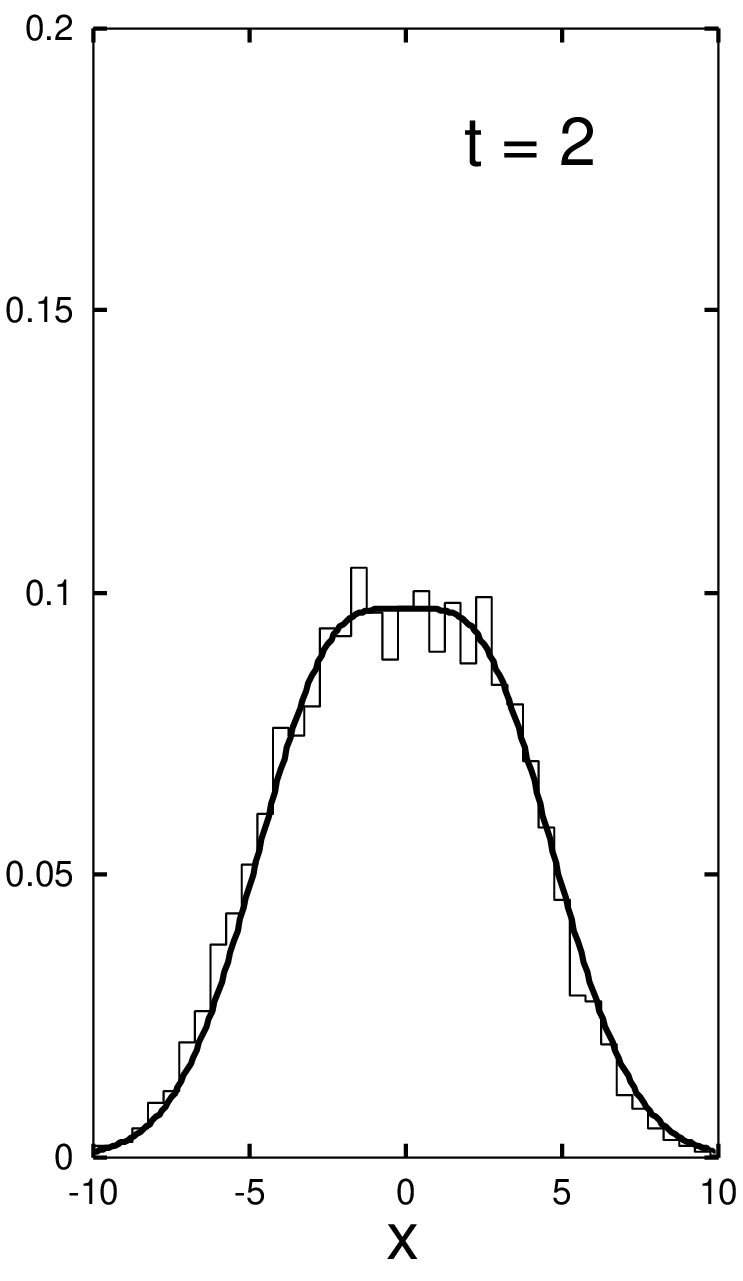}
\includegraphics[width=0.23\linewidth,height=0.43\linewidth]{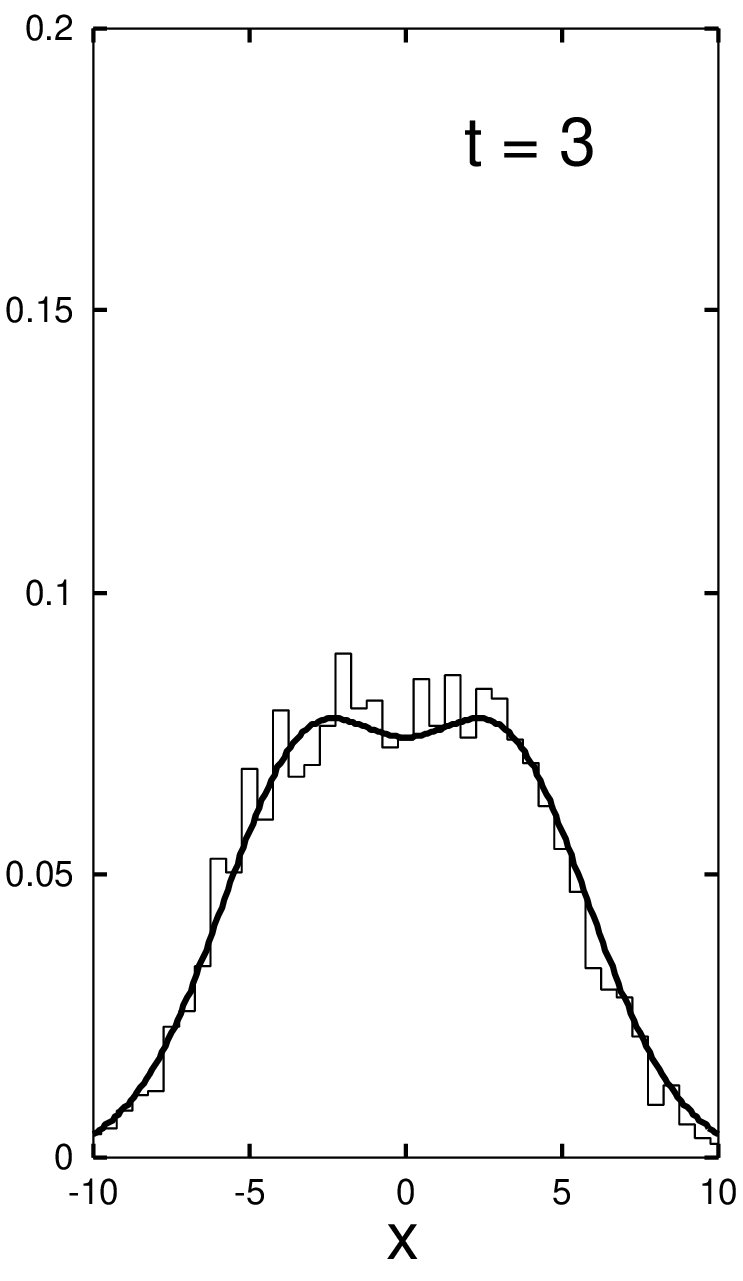}
\includegraphics[width=0.23\linewidth,height=0.43\linewidth]{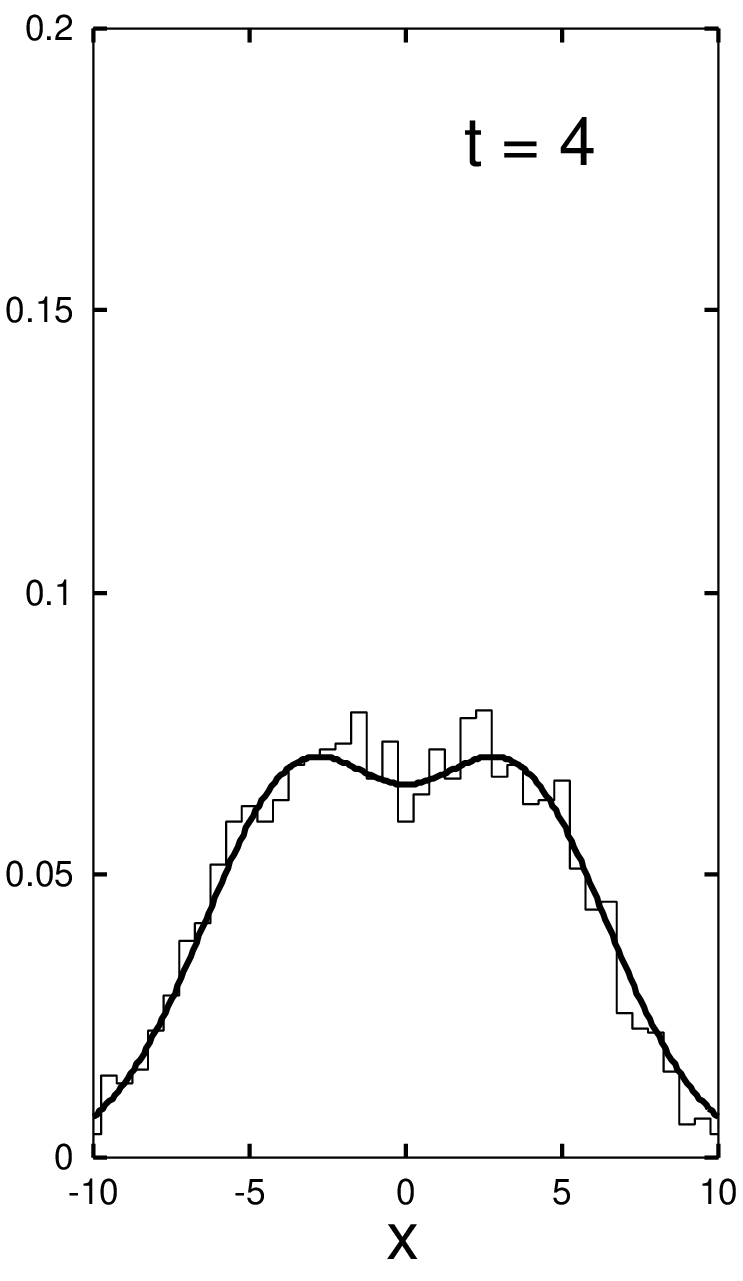}
\end{center}
\nsp
\caption{
The student field distribution
$P_{t}(x)$ generated during on-line
Hebbian learning, from
a teacher with a reversed wedge
of width $a=1$ and for $\eta=1$, $\alpha=0.5$ and $\gamma=0.5$, at times $t\in\{1,2,3,4\}$.
Solid lines: the theoretical result (\ref{Pt}). Histograms: results obtained via computer simulations
 for systems of size $N=10000$.
}
\label{fig2}
\end{figure}
It follows from \eref{Pt} that $P_t(x)$ is a symmetric
function of $x$,
for all times and all values of the
reversed wedge width $a$.
In the special cases $a=\infty$ and $a=0$ (where the task becomes realizable)
we find our result (\ref{Pt}) reducing to that of $\cite{Rae}$:
\begin{eqnarray*}
P_{t}(x) =
\int \frac{d\hat{x}}{2\pi}
{\rm e}^{-\frac{1}{2}Q\hat{x}^2 +
\chi_{\rm r}(\hat{x})}
\cos(\hat{x}x)\left\{
{\cos}(\chi_{i}(\hat{x}))
+(1\!-2\lambda)
G(\hat{x}R)
\sin (\chi_{\rm i}(\hat{x}))
\right\}
\end{eqnarray*}
with $\lambda=0$ for $a=\infty$, and $\lambda=1$ for $a=0$.
This is consistent with \cite{Rae},
where the parameter $\lambda$
denoted the probability that a teacher output was corrupted by noise.
Here we find that,
if the width of the reversed wedge is
zero, the transfer function of
the teacher is the inverse of that of the student, and the  the output of
the teacher can be regarded as a noisy output
with flip probability $\lambda=1$.
In contrast, in the general case
$0<a<\infty$, equation (\ref{Pt}) shows that the effect of
structural non-realizability can not be described by
an `effective' output noise.
In \Fref{fig2} we plot
$P_{t}(x)$ as given by equation (\ref{Pt}), for
$a=1$, $\eta=1$ and $\gamma=0.5$, at different points in time, and
we compare the result to the corresponding observations in
numerical simulations (histograms).
One clearly observes how $P_{t}(x)$
evolves from a Gaussian distribution at
$t=0$ to a manifestly non-Gaussian one.

\begin{figure}
\begin{center}
\includegraphics[width=0.44\linewidth]{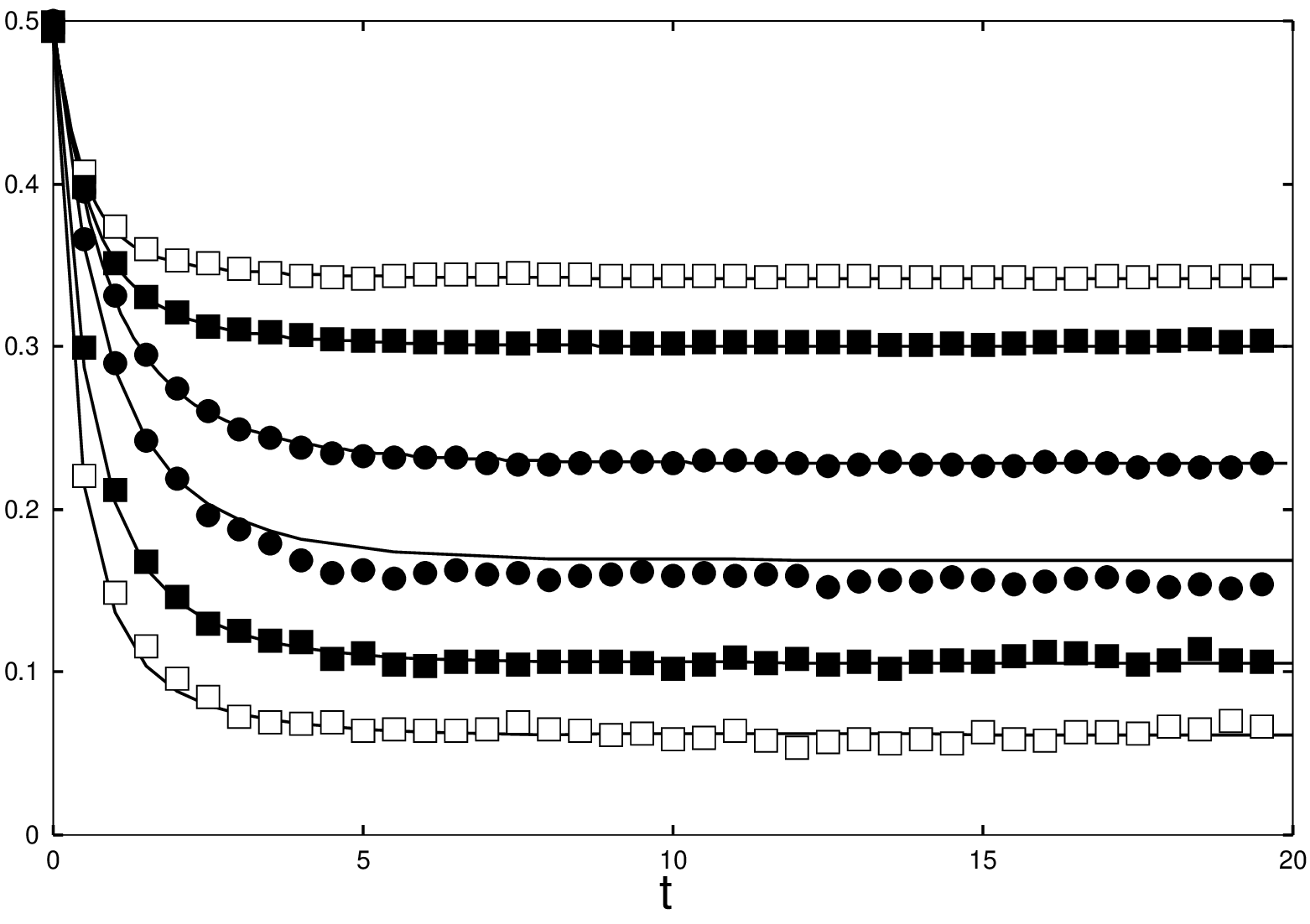}\hspace{0.2cm}
\includegraphics[width=0.44\linewidth]{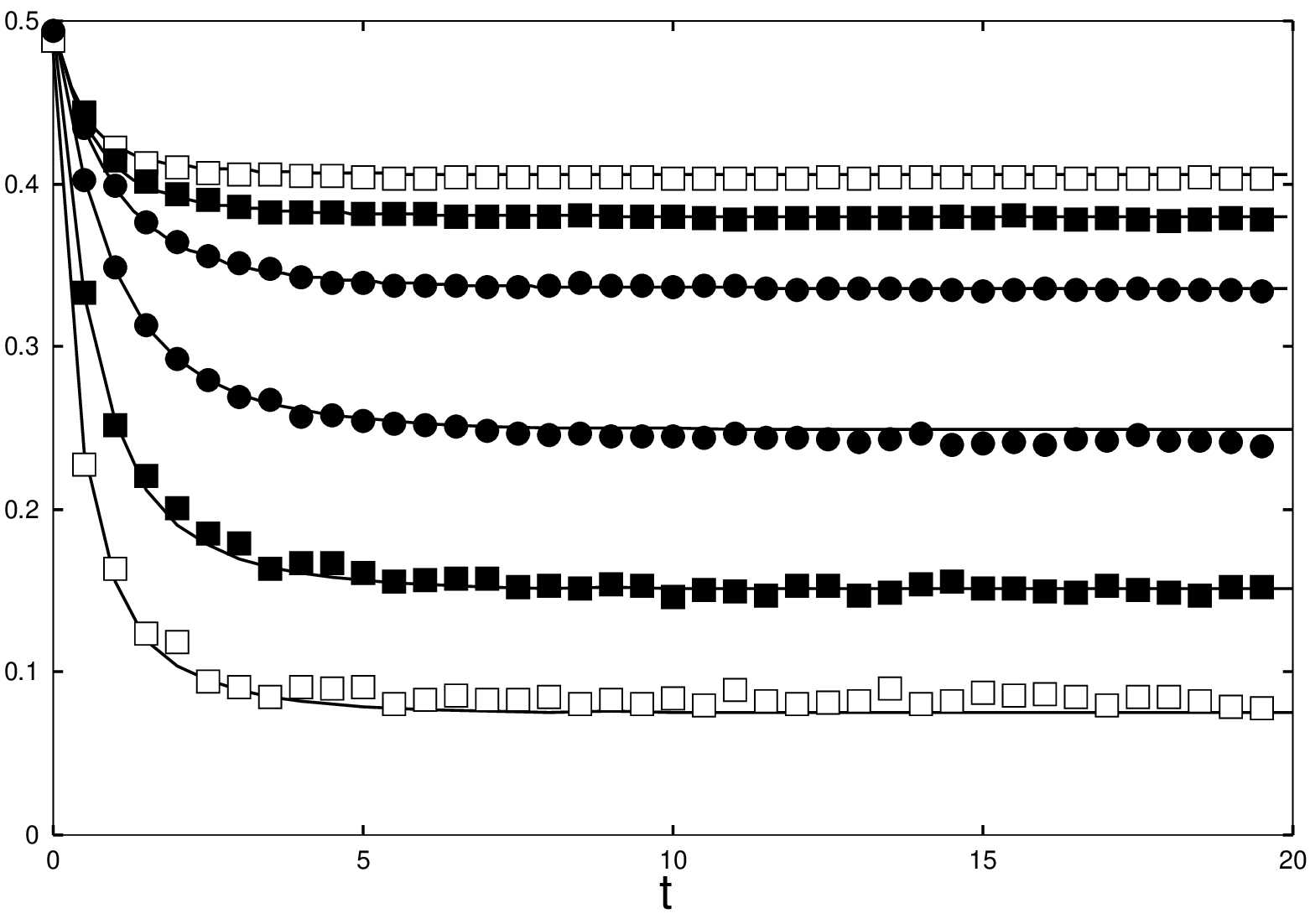}
\vspace{0.4cm}
\includegraphics[width=0.44\linewidth]{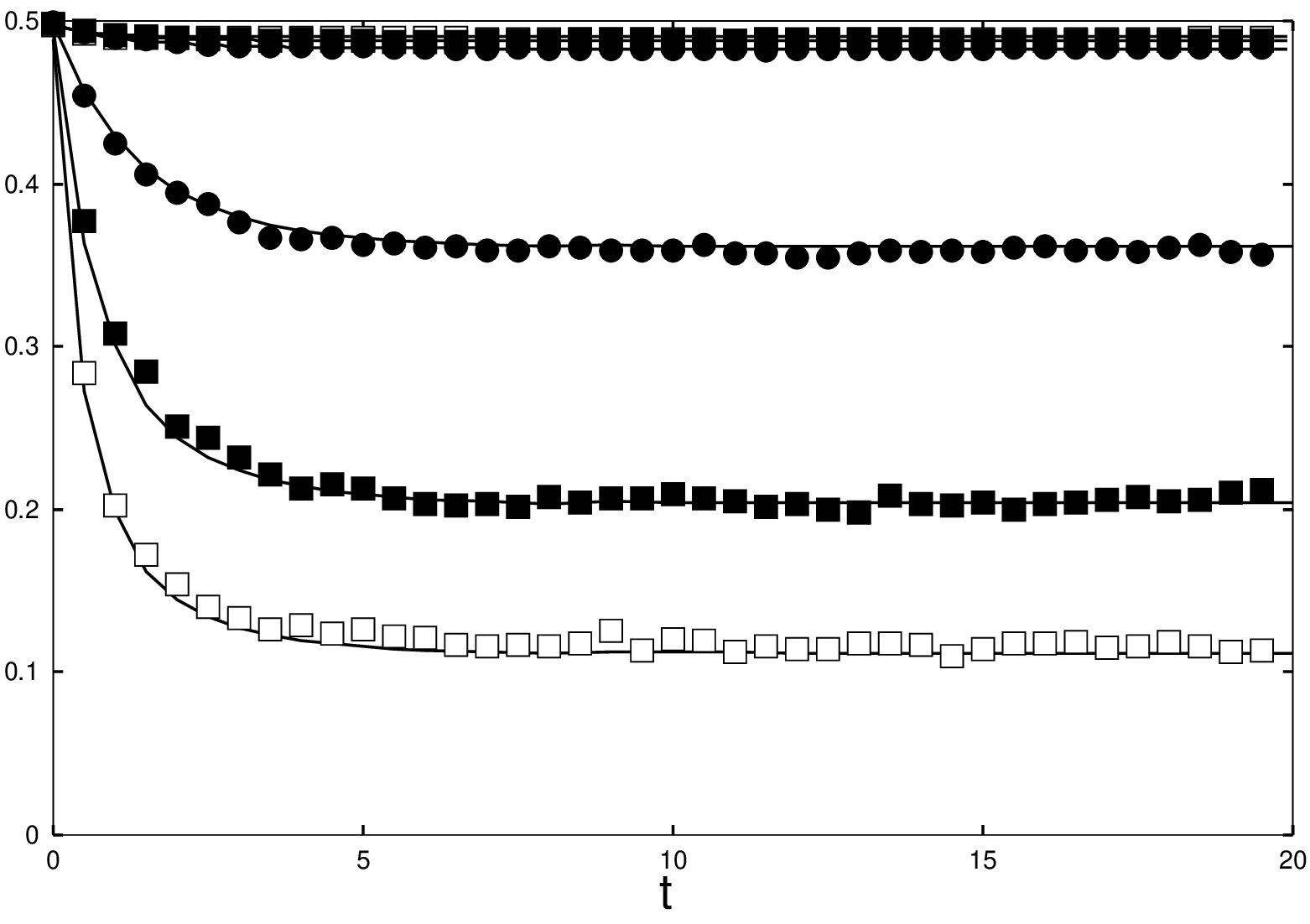}\hspace{0.2cm}
\includegraphics[width=0.44\linewidth]{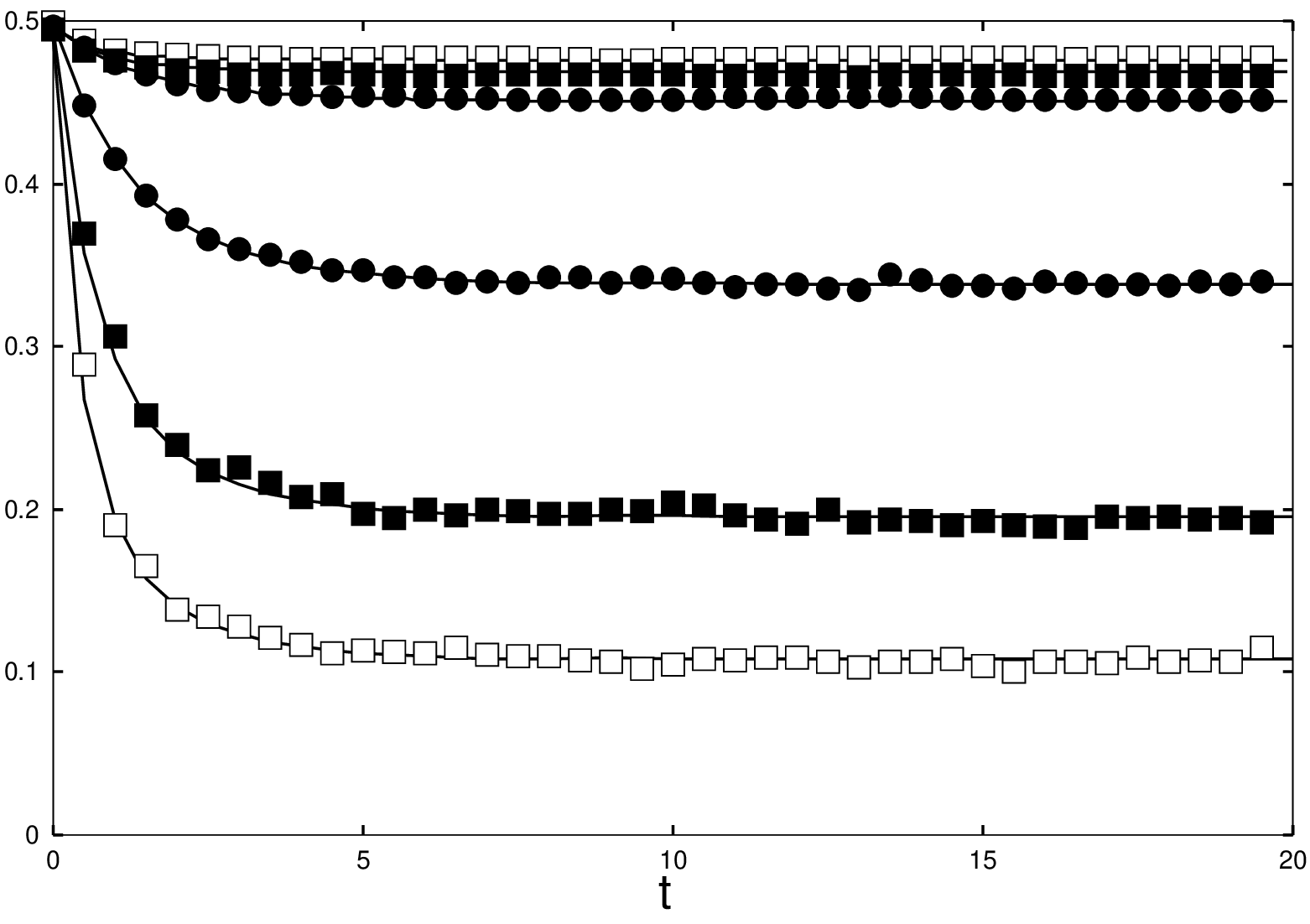}
\end{center}
\vspace*{-6mm}
\caption{
Training errors $E_t$ and generalization errors $E_g$ as functions of time, for different values of $\alpha$.
In all cases $\eta=1$ and $\gamma=0.5$, with initial conditions
$Q_{0}=1$ and $E_{g}(t=0)=0.5$.
From the upper left panel to
the lower right panel: $a=0.0,~0.5,~1.0$ and $1.5$.
In each panel, the upper three solid lines indicate our theoretical
results of $E_{g}$, together with the corresponding results of
computer simulations: \opensquare ($\alpha=0.5$),
\fullsquare ($\alpha=1.0$)  and  \fullcircle ($\alpha=4.0$).
The lower three lines are theoretical
results for $E_{t}$, compared to the results of computer simulations, with \opensquare ($\alpha=0.5$),
\fullsquare ($\alpha=1.0$)  and  \fullcircle ($\alpha=4.0$).
All simulations are carried out for systems
of size $N=5000$.}
\label{fig3}
\end{figure}

Finally, we calculate
the training error $E_{t}$, which measures the average fraction of errors
made by the student on inputs taken from the training set.
It is given by
\begin{eqnarray*}
E_{t} =
\int \!\!\! \int\! dx Dy \sum_{T_a = \pm 1}
{\Theta}[-T_a (y)S(x)] P_{t}(x|y,T_a)P(T_a|y).
\end{eqnarray*}
By using  \eref{relationP} and \eref{Pt} we can obtain the
explicit form of $E_{t}$ as
\begin{eqnarray}
E_{t}&=&\frac{1}{2}
-\int\! \frac{d\hat{x}}{2\pi \hat{x}}
{\rm e}^{-\frac{Q}{2}\hat{x}^{2}
+\chi_{\rm r}(\hat{x})}
[G(\hat{x}R) \cos (\chi_{\rm i}(\hat{x}))
-\sin (\chi_{\rm i}(\hat{x}))] \nonumber \\
&& +
4 {\rm e}^{-\frac{a^2}{2}}
\int\! \frac{d\hat{x}}{2\pi \hat{x}}
{\rm e}^{-\frac{Q}{2}\hat{x}^{2}
+\chi_{\rm r}(\hat{x})}
{\cos}(\chi_{\rm i}(\hat{x}))
\int_{0}^{\hat{x}R}\!\!
\frac{d\lambda}{\sqrt{2\pi}}
{\rm e}^{\frac{\lambda^2}{2}}
\cos (a\lambda).
\label{eq:Et}
\end{eqnarray}
In \Fref{fig3}, we
plot both the training error (\ref{eq:Et}) and the generalization error
(\ref{gene}) for four different values of
the width $a$ of the teacher's reversed wedge, viz. $a\in\{0.0,0.5,1.0,1.5\}$.
In all case we find the theoretical results and the
computer simulations to be in excellent agreement.
In the limit $t \rightarrow \infty$ we also observe that the
asymptotic values of both $E_{g}$ and $E_{t}$
indeed approach $E_{g}^{*}$ (see \eref{asymge}) for increasing
 $\alpha$, as it should.

In conclusion, in this letter we have solved the dynamics of on-line
Hebbian learning with structurally unrealizable restricted training sets
exactly, for the case where a standard perceptron is being trained
by a teacher perceptron with a reversed wedge transfer function.
Although our solution applies only to
Hebbian learning (as did the one in \cite{Rae}),
we believe that our results provide a valuable new benchmark
against which to test (approximations made in) more general formalisms,
such as generating functional analysis \cite{Horner1,Horner2,HC},
dynamical replica theory \cite{CS1,CS2} or the cavity method \cite{WongLi}.
\\

\noindent
The authors would like to thank King's College London (JI) and
the Tokyo Institute of Technology (ACCC) for their hospitality.

\section*{References}

\end{document}